\documentclass[journal,twoside,web]{ieeecolor}
\usepackage{generic}
\usepackage{cite}
\usepackage{amsmath,amssymb,amsfonts}
\usepackage{graphicx}
\usepackage{textcomp}
\usepackage{epsfig}
\usepackage{amsmath}
\usepackage{amssymb}
\usepackage{bm}
\usepackage{amsfonts}
\usepackage{graphics}
\usepackage{subfigure}
\usepackage{algorithm}
\usepackage{algpseudocode}
\usepackage{color}
\usepackage{tikz,pgfplots,pgfplotstable}
\allowdisplaybreaks
\def\BibTeX{{\rm B\kern-.05em{\sc i\kern-.025em b}\kern-.08em
    T\kern-.1667em\lower.7ex\hbox{E}\kern-.125emX}}
\algnewcommand{\Initialize}[1]{
  \State \textbf{Initialize:}
  \Statex \hspace*{\algorithmicindent}\parbox[t]{.8\linewidth}{\raggedright #1}
}
\newtheorem{theorem}{Theorem}
\newtheorem{lemma}{Lemma}

\newtheorem{remark}{Remark}

\newtheorem{defn}{Definition}
\newtheorem{assum}{Assumption}

\begin{document}
\title{Completely Uncoupled User Association Algorithms 
 for State Dependent Networks}

\author{S.~Ramakrishnan, Venkatesh Ramaiyan, and K.~P.~Naveen
\thanks{S.~Ramakrishnan, and Venkatesh Ramaiyan are with IIT Madras, Chennai, 600036, 
India (email: ee12d036@ee.iitm.ac.in, rvenkat@ee.iitm.ac.in).
K.~P.~Naveen is with IIT Tirupati, Tirupati, 517506, India (email: naveenkp@iittp.ac.in). }
\thanks{The conference version of the work has appeared in 2017, IEEE Wireless Communication and Networking Conference (WCNC) \cite{7925713} }}
\maketitle

\begin{abstract}
We study a distributed user association algorithm for a heterogeneous 
wireless network with the objective of maximizing the sum of the 
utilities (on the received throughput) of wireless users. We consider
a state-dependent wireless network, where the rate achieved by the
users are a function of their user associations as well as the state of
the system. We consider four different scenarios depending on the
state evolution and the users' knowledge of the system state.  
In this context, we present completely uncoupled user association 
algorithms for utility maximization, where the users' association is 
entirely a function of its past associations and its received throughput. 
In particular, the user is oblivious to the association of the other users 
in the network. Using the theory of perturbed Markov chains, 
we show the optimality of our algorithms under appropriate scenarios.
\end{abstract}


 \section{Introduction}
In the present wireless scenario, a mobile user has the 
luxury to choose between several Access Points (APs), that are possibly enabled with 
different access technologies (e.g., WiFi, WiMAX, LTE, etc). The APs could be further 
heterogeneous in terms of their size (e.g., cellular, femto-cell, WiFi AP), and could be 
owned by different service providers. Thus, it may not be possible to expect a 
centralized coordination among different APs. 

In the above context, we are interested in designing distributed user association 
algorithms with the objective of optimizing system utilization. A key challenge in such 
design is the unavailability of information with each user regarding other users' 
behaviour (in terms of their association and the utilities they receive). Further, due to 
privacy concerns (since different service providers are involved), the APs may be 
reluctant to share some system level parameters (e.g., their transmit 
powers, pricing strategy, etc.) with the users. Thus, user association algorithms are 
expected to be \emph{completely uncoupled} \cite{marden-etal12pareto-optimality}, 
whereby a user's association-decision is entirely based on its past decisions and the 
utilities it received in the past. In this paper, we design such a completely uncoupled 
user association algorithm for a state based system, comprising of finite number of 
states, BSs, and users.
\par In a state based network, the throughput or pay-off received by a user depends on the 
system state, in addition to the users' association choices. The state could represent
any background process, which the users do not have any control over.
The following examples illustrate the need for state based model considered in this paper.
\begin{enumerate}
\item Delayed pay-off: The pay-off received by a user is delayed by a fixed unknown 
time. In this case, the state represents a moving window of previous associations.
\item Network Response: In the user association problem, the base station could 
employ channel selection or power control algorithms. These algorithms could be 
a function of the association choices. Here, the state is modelled as an independent random 
process depending on the action. In a more interesting case, these algorithms could depend on the 
previous state in addition to the current association choices. 
In this case, the state is modelled as a controlled Markov process, 
where the control action correspond to the association choices. 
\item Wireless Channel: The state could represent fading in wireless channel. In 
typical wireless network, fading is generally modelled as an ergodic random 
process \cite{1374962}.     
\end{enumerate}


\par In this paper, we consider four cases depending on user's knowledge of the 
state and state transition model. 
\par We start with the case where the state is unknown to the users.
When the state is unknown, we consider two state transition models. First, we assume that state transition 
is a deterministic function of the prior state and the current association vector. 
Second, we consider the case where the state is an $iid$ random variable depending on the association vector chosen.
We then consider the case where the state is known to users.
For the case where state is known, we first consider a more general controlled Markov state transition model.
Finally, we consider the case with ergodic state transitions. 
A formal description of the considered system model will be presented in 
Section~\ref{system_model_section}. Before proceeding further, we end 
this section with a brief survey of related literature.

\textbf{\emph{Related Work:}}
Utility maximization is known to achieve notions of fairness \cite{Kelly1998}. For example, 
$\log$ utility is known to achieve proportional fairness \cite{Kelly1998}. In \cite{1310314},
Kushner and Whiting showed the convergence of gradient algorithms in a time varying environment.
The above algorithms are centralized in nature i.e., they require information about all the nodes in the network.

\par In \cite{DBLP:journals/ton/JiangW10}, Jiang and Walrand proposed CSMA based distributed scheduling 
algorithms for a conflict graph model, for which proof of optimality was shown in \cite{5625654}.
In \cite{4215753} Kauffmann {\it et al}., proposed distributed channel selection and user association
algorithms for IEEE 802.11 networks using Gibbs sampler for some tailored utilities. 
In \cite{borst2014nonconcave}, Borst {\it et al}., showed that  maximizing utilities using Gibbs sampling
require two hop information, if the utility depends on one hop neighbours.   

\par Uncoupled learning algorithms were popularized by Young in the context of coordination games,
in his seminal work \cite{young93evolution-of-conventions}. Over the years, several variants 
of the algorithm in \cite{young93evolution-of-conventions} have been studied.
For instance, Pradelski and Young \cite{pradelski-young12nash-equilibria} proposed an algorithm
for achieving efficient Nash equilibrium in general $n$ person games satisfying interdependence property,
while the problem of obtaining pareto optimal solution has been considered by Marden et al., in \cite{marden-etal12pareto-optimality}.
Also, in \cite{Borowski2018}, Borowski and Marden proposed a completely uncoupled algorithm for 
achieving efficient correlated equilibrium under interdependence assumption.
Algorithms for state space based potential games have been studied in \cite{marden12state-based}.
In contrast, we study utility maximization in state based networks satisfying interdependence property.  

In the context of wireless networks, algorithms for user association are available in the literature 
(see e.g., \cite{beherano-etal07fairness,coucheney-etal09user-association,haddad-etal11hybrid-approach}).
However, these are either centralized \cite{beherano-etal07fairness}, or require message passing within the
network \cite{coucheney-etal09user-association,haddad-etal11hybrid-approach}.
Singh and Chaporkar \cite{singh-chaporkar13user-association} were the first to design uncoupled user association
algorithm for wireless networks. Similar to the objective in \cite{marden-etal12pareto-optimality}, 
the authors in  \cite{singh-chaporkar13user-association} consider the problem of maximizing the sum of user 
payoffs. The algorithm in \cite{singh-chaporkar13user-association} is essentially based on the algorithm proposed 
in \cite{marden-etal12pareto-optimality}. Similarly, in our prior work \cite{7746656} we have adapted the algorithm 
in \cite{marden-etal12pareto-optimality} to obtain a distributed algorithm for maximizing the sum of user utilities.
However, in \cite{7746656} we assume that the utilities are a function of the long-term throughput achieved by the users,
rather than the instantaneous throughput as considered in \cite{singh-chaporkar13user-association}. 
%

%
In this work, we generalize the setting in \cite{7746656} by incorporating a state evolution into the model. 
To the best of our knowledge, the particular setting we consider, and the corresponding optimality result 
we obtain is not available in the literature.

\textbf{\emph{Paper Outline:}} 
In Section~\ref{system_model_section}, we formally discuss our system model. 
In Section~\ref{Section:Deterministic state evolution}, we propose distributed 
algorithm and show optimality under deterministic state evolution.
Then, we consider $ iid$ state evolution depending on the association vector in 
Section~\ref{Section:Unknown state and Random state transition}, where the 
users are oblivious to the system state.
Under complete state knowledge, we propose an optimal distributed user 
association algorithm, when the state evolution is
a controlled Markov process in Section \ref{Section:KnownStateControlledMarkov} and any ergodic process in 
Section~\ref{Section:Known state and Random state transition}. 

 \section{System Model}
\label{system_model_section}
We consider a wireless system comprising $M$ Access Points (APs) and $N$ users.
Let $\mathcal{M}$ and $\mathcal{N}$ denote the set of APs and users, respectively.
The APs could be heterogeneous in terms of their wireless technology 
(e.g., WiFi, WiMAX, LTE) and size (e.g., cellular, femto-cell, Wifi AP). We assume that
 each user can associate  with a subset of these APs. Such a limitation could arise,
possibly, because of the proximity of a user to only some APs, or due to the limited 
wireless technologies available on their user-equipments. Specifically, let 
$\mathcal{A}_i\subseteq\mathcal{M}$ be the subset of APs with which 
user $i\in\mathcal{N}$ can associate.

We assume a time slotted system. In time slot $t\in\mathbb{N}$,
user $i\in\mathcal{N}$ is associated with a single AP $a_i(t)$ where 
$a_i(t)\in\mathcal{A}_i$.
Let $\bm{a}(t):=(a_1(t),\ldots,a_N(t))$ denote the vector of associations of all users.
The set of all possible association-vectors is denoted as
$\mathcal{A}:=\mathcal{A}_1\times\cdots\times\mathcal{A}_N$.

It is usually assumed that the rate achieved by a user in a given time-slot
is a function of the vector of associations in that slot 
(see e.g., \cite{singh-chaporkar13user-association,7746656}).
In our work, we generalize the above setting by introducing a finite set, 
$\mathcal{S}$, of system states, and assume that the users' rate is a function 
of the system state and the association vector in the current time slot.
Thus, if $s(t)\in\mathcal{S}$ is the system state at time $t$, then the rate, $r_i(t)$,
achieved by user-$i$ in slot $t$ is given by,
\begin{eqnarray}
\label{rate_equation}
 r_i(t)= f_i(s(t),\bm{a}(t)),
\end{eqnarray}
where, $f_i:\mathcal{S}\times\mathcal{A}\rightarrow\Re_+$, for all $i\in\mathcal{N}$. 
Without loss of generality, we assume that $r_i(t)$ lies between $0$ and $1$.

\textbf{\emph{Average Rate and Utilities:}}
Let $s(0)=s_0$ be the initial state of the system. Then, given a sequence of 
association vectors 
$\{ \bm{a}(t) : t \in {\mathbb N} \}$, the long-term average
rate received by user $i$ can be written as,
\begin{eqnarray}
\label{average_rate_equation}
 \overline{r}_i(s_0, \{\bm{a}(t)\}) = \liminf_{T\rightarrow\infty} \frac{1}{T} 
 \sum_{t=0}^{T-1} f_i(s(t),\bm{a}(t))
\end{eqnarray}
 Different sequences of association vectors can yield possibly different long-term rate vectors.
Let ${\mathcal R}(s_0)$ denote the set of all
such feasible long-term average rate vectors
 $\{ (\overline{r}_1, \cdots, \overline{r}_N) \}$.

\par The utility achieved by user-$i$ is measured using a utility function $U_i$, which
is a function of the average rate, $\overline{r}_i$. 
We assume that the utility functions are continuous and satisfy, for all $i\in\mathcal{N}$,
$0\le U_i(\overline{r}_i) \le u_{\max} <1 $ for all possible $\overline{r}_i\in[0,1]$ 
(i.e., the utility functions are continuous and bounded, but are general otherwise).

\textbf{\emph{Optimization Framework:}} We are interested in maximizing the sum of utilities
of all users. Formally, we consider:
\begin{align}
\left.
\begin{aligned}
\label{optimization_problem}
 \underset{\{\bm{a}(t):t \in {\mathbb N} \}}{\mbox{Maximize: }} & \sum_{i\in\mathcal{N}} U_i(\overline{r}_i)  \\
 \mbox{Subject to: } & s(0) = s_0, \mbox{ and } \forall i\in\mathcal{N}, t\ge0 \\
 &  s(t) \in \mathcal{S} \\
 & r_i(t) = f_i(s(t),\bm{a}(t)) \\
 & \overline{r}_i = \liminf_{T\rightarrow\infty} \frac{1}{T} \sum_{t=0}^{T-1} r_i(t).
 \end{aligned}
 \right\}
\end{align}


The problem in \eqref{optimization_problem} can be solved in a centralized 
manner, assuming that the state is known before the association decision is 
made \cite{1374962}. Our goal is to solve \eqref{optimization_problem}
 in a completely distributed manner. In the following sections, 
 we shall consider four scenarios depending on the state evolution and 
 user's knowledge of the states. 

\par First we shall assume that the users do not know the system state.
Under this assumption, we shall consider two cases: (i) State 
evolves deterministically and (ii) the State depends only on the action and evolves $iid$ 
over time. For deterministic state evolution, we will propose a 
distributed algorithm with stationary performance close to the optimal utility. 
Also, this optimal sum utility is no worse compared to the centralized solution 
with complete state knowledge. For the $iid$ case, we propose an algorithm that maximizes the sum utility, 
where the utility is a function of the expected average rate (w.r.t state).    
This is optimal under the assumption that the users do not know the system state.

\par Next, we shall assume that the users have complete knowledge of the state.
Under this assumption, we shall consider two cases: (i) State evolves as
a controlled Markov process and (ii) State evolves as an Ergodic random process.
For the controlled Markov evolution, we propose a distributed algorithm that maximizes the sum utility,
where the utility is a function of the expected average rate (w.r.t the stationary distribution of the state).   
For the ergodic evolution, we propose a distributed utility maximization algorithm that is optimal. 

 \section{Unknown State: Deterministic evolution}
\label{Section:Deterministic state evolution}
In this section, we assume that the system state evolves deterministically 
and is unknown to the users. Formally, the state transition is as follows, 
\begin{eqnarray}
\label{transition_equation}
 s(t+1) = g(s(t),\bm{a}(t)),
\end{eqnarray}
where, $g:\mathcal{S}\times\mathcal{A}\rightarrow\mathcal{S}$.
The rate functions ($f_i(\cdot)$) and the state transitions ($g(\cdot)$) are 
deterministic functions of the current state and the current association-vector.
We make the following irreducibility assumption about the wireless system.
\begin{assum}
(Irreducibility)
\label{assum:Irreducibility}
Given any pair of states $s,s'\in\mathcal{S}$, there exists a finite sequence
of association vectors $(\bm{a}^{(1)}, \ldots, \bm{a}^{(n)})$ such that 
$s^{(1)}=s$, $s^{(j+1)}=g(s^{(j)},\bm{a}^{(j)})$ for all $j=1,\ldots,n$, and $s^{(n+1)}=s'$.
\end{assum}
The above assumption insists that all the states can be visited by choosing 
an appropriate sequence of association vectors. Thus, it follows that the set
 of feasible long-term average rates in \eqref{average_rate_equation} is independent
 of the initial state of the system, i.e., ${\mathcal R}(s_0) = {\mathcal R}$.
The set ${\mathcal R}$ is usually referred to as the \emph{rate region}
 of the wireless system.
The formulation in \eqref{optimization_problem} could be re-written as, 
\begin{align}
\label{rate_region_problem}
\underset{(\overline{r}_1,\ldots,\overline{r}_N) \in {\mathcal R}}{\mbox{Maximize:}} 
& \ \ \  \sum_{i \in {\mathcal N}} U_i(\overline{r}_i).
\end{align}

\par The formulation in \eqref{rate_region_problem} requires us to seek an optimal
sequence of association-vectors from the set of all possible infinite length sequences. 
We simplify this formulation in \eqref{rate_region_problem} 
(in Section~\ref{cycles_section}) before 
proceeding to the design and analysis of an optimal user association algorithm 
(Sections~\ref{Subsection:Deterministic Algorithm} and \ref{optimality_section}).



\subsection{Configuration Cycles}
\label{cycles_section}
{In this section, we identify cycles of state and association-vector pairs such that 
convex combinations of these cycles can achieve any rate vector in the rate region 
$\mathcal{R}$. This representation will enable us to simplify the formulation in 
(\ref{rate_region_problem}). We begin with the following definitions.}

A pair $(s,\bm{a})$ of state $s\in\mathcal{S}$,
and association-vector $\bm{a}\in\mathcal{A}$, is referred to as a \emph{configuration}.
A sequence of configurations, $\bm{c}=(c^{(1)},\ldots,c^{(n)})$, where $c^{(j)}=(s^{(j)},a^{(j)})$ for $j=1,2,\ldots,n$,
is said to be a \emph{configuration cycle} (or simply cycle) if it satisfies:
 $s^{(j+1)}= g(c^{(j)})$ for all $j=1,\ldots,n-1$,
and $g(c^{(n)}) = s^{(1)}$. If the sequence of configurations are distinct, i.e.,
if $c^{(1)}, \ldots, c^{(n)}$ are distinct elements, then the configuration cycle is called a 
\emph{basic} configuration cycle. Clearly, the length of any basic configuration cycle
is restricted to be not more than $|{\mathcal S} \times {\mathcal A}|$.

Let ${\mathcal{C}}$ denote the set of all basic configuration cycles.
Clearly, the set ${\mathcal C}$ is non-empty and finite.
%
%
Given a basic configuration cycle $\bm{c} \in {\mathcal{C}}$ of length $|\bm{c}|$,
the average rate achieved in the cycle $\bm{c}$ by user-$i$, denoted $r_i(\bm{c})$,
is defined as
\begin{eqnarray}
 {r}_i(\bm{c}):= \frac{1}{|\bm{c}|}\sum_{j=1}^{|\bm{c}|} f_i(c^{(j)})
\end{eqnarray}
Let $({r}_1(\bm{c}), \ldots,  {r}_N(\bm{c}))$ denote the vector of user 
rates achievable in the cycle $\bm{c}$. 
The following lemma relates the set of rates achievable using the basic
configuration cycles and the rate region ${\mathcal R}$ of the wireless system.
\begin{lemma}
Let ${\mathcal R}_{\mathcal C}$ denote the convex hull of the rate vectors 
achievable using the basic configuration cycles, i.e.,
 \begin{align*}
  {\mathcal R}_{\mathcal C} = \left\lbrace (\overline{r}_1, \ldots, \overline{r}_N) \;\middle|\;
 \begin{aligned}
 & \overline{r}_i = \sum_{\bm{c} \in {\mathcal C}} p_{\bm{c}} r_i(\bm{c}),\\ 
 &p_{\bm{c}} \geq 0 \;\; \forall \bm{c} \in {\mathcal C},  
  \sum_{\bm{c} \in {\mathcal C}} p_{\bm{c}} \leq 1 
 \end{aligned}
 \right \rbrace
 \end{align*}
Then, ${\mathcal R}_{\mathcal C} = {\mathcal R}$.
\hfill \QED
\end{lemma}

The above lemma permits us to propose an equivalent formulation of the 
optimization problem (\ref{rate_region_problem}) in terms of the basic 
configuration cycles.
\begin{align}
\left.
\begin{aligned}
\label{equivalent_problem}
{\mbox{Maximize: }} & \sum_{i\in\mathcal{N}} U_i(\overline{r}_i) \\
{\mbox{Subject to:}} \ & \overline{r}_i = \sum_{\bm{c}\in{\mathcal{C}}} p_{\bm{c}} r_i(\bm{c}) \\
 & p_{\bm{c}} \ge 0, \sum_{\bm{c} \in {\mathcal C}} p_{\bm{c}} \leq 1
\end{aligned}
\right\}
\end{align}
In the subsequent section, we will discuss a user association algorithm that can
achieve any time average of the basic configuration cycles that optimises the above 
formulation.

 \subsection{User Association Algorithm}
\label{Subsection:Deterministic Algorithm}
In this section, we present a \emph{completely uncoupled} user association 
algorithm for a state dependent wireless network, where the state transitions are deterministic.
{In a {completely uncoupled} scenario, a user can observe
only its past actions and received utilities; the actions and utilities
 of other users are not known. In fact, a user can be completely oblivious
  about the existence of other users in the system. In this case, we will further 
assume that the users cannot observe the system state as well.}
We note that the algorithm presented here generalizes the techniques
studied in work such as \cite{marden-etal12pareto-optimality} and \cite{7746656}.

\begin{algorithm}
\caption{\textbf{: User Association Algorithm}}
\label{Alg1: Deterministic case}
\begin{algorithmic}
\Initialize{Fix $z > N$, $K_{max} \in \mathbb{Z}^+$ and $\epsilon > 0$. \\ For all $i \in {\mathcal N}$, set $K_i(0)$ uniformly from $\{ 1, \cdots, K_{max} \}$. \\ For all $i \in {\mathcal N}$, set $q_i(0) =0$.}
\end{algorithmic}

\begin{algorithmic}
\State \textbf{Update for user association at time $t$:}
\If {($q_i(t-1)=1$)}
\State
$a_i(t) =
\begin{cases}
a_i(t-K_i(t - 1)) & \text{w.p. \ } 1 - \epsilon^z \\
a_i \in A_i & \text{w.p. \ } \frac{ \epsilon^z}{\left|A_i\right|}
\end{cases}$
\Else
\State $a_{i}(t)= a_i \in A_i$ w.p. $\frac{ 1}{\left|A_i\right|}$
\EndIf
\end{algorithmic}

\begin{algorithmic}
\State \textbf{Update for $q_i(\cdot)$ and $K_i(\cdot)$ at time $t$:}
\If { ($q_i(t-1)=1$) and ($a_i(t)=a_i(t-K_i(t-1))$) \\ and $\left( r_i(t) = r_i(t-K_i(t-1))\right)$}
\State $K_i(t) = K_i(t-1)$
\State $q_i(t) = 1$
\Else
\State Pick $K_i(t)$ uniformly from $\{ 1, \cdots, K_{max}\}$
\State
$q_i(t) =
\begin{cases}
1 & \text{w.p. \ } \epsilon^{1-U_ i\left(\frac{1}{K_i(t)}\sum\limits_{j=t-K_i(t)+1}^{t}  r_i(j)\right)} \\
0 & \text{w.p. \ } 1- \epsilon^{1-U_ i\left(\frac{1}{K_i(t)}\sum\limits_{j=t-K_i(t)+1}^{t}  r_i(j)\right)}
\end{cases}$
\EndIf
\end{algorithmic}
\end{algorithm}

Our user association algorithm is presented in Algorithm~1.
In the following, we describe the working principles of our algorithm. 
Suppose,  at every time $t-1$, each user $i\in\mathcal{N}$ maintains its past
associations  $(a_i(1), \cdots, a_i(t-1))$ and throughputs received
$(r_i(1), \cdots, r_i(t-1))$. Further, let the users maintain an 
internal ``satisfaction'' variable $q_i(t-1)$, and an averaging window size $K_i(t-1)$.
We let $q_i(\cdot)$ take values from the binary set $\{ 0,1 \}$, where $q_i(\cdot) = 1$
represents a state of ``content'' with the choice of user association and the average 
throughput received (in the previous $K_i(\cdot)$ slots), while
$q_i(\cdot) = 0$ represents a state of ``discontent'' for the user.
The averaging window size, $K_i(\cdot)$, is used to average the received 
throughput (and also identifies the sequence length of actions that are repeated) 
with $K_i(\cdot)$ taking values from the set $\{1, \cdots, K_{max}\}$ 
where $K_{max}\in\mathbb{N}$ is fixed.

The choice of user association, $a_i(t)$, made at the beginning of the slot $t$ 
is entirely a function of the internal satisfaction variable $q_i(t-1)$.
When a user $i$ is content at the beginning of slot $t$,
i.e., when $q_i(t-1) = 1$, the user repeats an earlier action, here $a_i(t-K_i(t-1))$,
with high probability $1 - \epsilon^z$ 
(where $z$ is a parameter satisfying $z > N$, the number of users). When a user
$i$ is discontent at the beginning of slot $t$, i.e., when $q_i(t-1) = 0$, 
then the user selects an association uniformly from ${\mathcal A}_i$.

The internal satisfaction variable $q_i(t)$ and the averaging window size 
$K_i(t)$ are updated at the end of slot $t$.
If the user $i$ was content in slot $t-1$ (i.e., when $q_i(t-1) = 1$), then, 
the user continues to remain content in slot $t$
if it had repeated an earlier action (i.e., if $a_i(t) = a_i(t-K_i(t-1))$, 
which happens with high probability)
and if it had received the same throughput as in the slot 
$t - K_i(t-1)$, i.e., $r_i(t) = r_i(t-K_i(t-1))$
(which could happen when the vector of user associations and the 
system state remains unchanged).
Otherwise, a user becomes content with a very low probability depending on
the utility~($U_i$) of the average throughput received by the user in the previous 
$K_i(t-1)$ time slots. 

We note that when all the users are content, i.e., when $q_i(\cdot) = 1$
for all $i$, then, the users repeat their last $K_i(\cdot)$ actions 
(in synchrony) and continue to receive a constant average throughput 
(if the corresponding action sequence of  length $K_i$ is a configuration cycle)
based on the sequence of actions and the system states. 
A sequence of actions is preferred depending on the average 
user throughput corresponding to the $K_i(\cdot)$ association sequence 
and the user utilities, and is a function of $\epsilon$ as well.
In the following section, we will show that Algorithm~1 chooses an action sequence
that optimises the formulation in (\ref{equivalent_problem})  
as $\epsilon \rightarrow 0$ and as $K_{max} \rightarrow \infty$.

 \subsection{Optimality Results}
\label{optimality_section}
In this section, we show that Algorithm~1 selects a sequence of associations 
for users that tends to optimize the formulation in (\ref{equivalent_problem}).
Define $X_{\epsilon}(t)$ as 
\[ X_{\epsilon}(t)=(c(t-K_{max}+1),\dots,c(t),\bm{K}(t),\bm{q}(t)) \]
where $c(t) =  (s(t), \bm{a}(t))$,
$\bm{K}(t) = (K_1(t), \cdots, K_N(t))$ and $\bm{q}(t) = (q_1(t), \cdots, q_N(t))$.
Let ${\mathcal K}$ and ${\mathcal Q}$ denote the state space of $\bm{K}(t)$ and $\bm{q}(t)$.
$X_{\epsilon}(t)$ corresponds to the recent configuration states of the system, 
the vector of averaging window sizes and the satisfaction variables 
of the $N$ users in the current slot $t$.
The following lemma shows that the random process 
$\{ X_{\epsilon}(t) : t \in {\mathbb N}\}$ 
is a regular perturbed Markov chain (perturbed by the algorithm 
parameter $\epsilon$) with a positive stationary distribution.

\begin{defn} $\{ X_\epsilon(t) \}$ is a regular perturbed Markov process 
(perturbed by $\epsilon$) if the following conditions
are satisfied (see \cite{young93evolution-of-conventions}).
\begin{enumerate}
\item $\forall \epsilon > 0$, $\{ X_{\epsilon}(t) \}$ is an ergodic Markov Process
\item $\begin{aligned}[t]
\forall \omega,\omega^{\prime} \in \Omega, \lim_{\epsilon \to 0} P^{\epsilon} \{ \omega , \omega^{\prime} \} = P^{0}\{\omega ,\omega^{\prime}\}
\end{aligned}$
\item $ \begin{aligned}[t]
\forall \omega,\omega^{\prime} \in \Omega,\ \text{if\ } &P^{\epsilon} \{ \omega , \omega^{\prime}\} > 0 \text{ for some } \epsilon > 0, \text{then},
\end{aligned}$
\[ 0< \lim_{\epsilon \to 0} \frac{ P^{\epsilon} \{ \omega , \omega^{\prime}\}}{\epsilon^{r(\omega , \omega^{\prime})}} < \infty \]
for some $r(\omega , \omega^{\prime})\geq 0$ and $r(\omega, \omega^{\prime})$ is called the resistance of the one-step transition $\omega , \omega^{\prime}$.
\end{enumerate}
\end{defn}

 \begin{lemma}
 \label{Regular perturbed Markov chain}
$\{ X_{\epsilon}(t) \}$ induced by Algorithm~1 is a regular perturbed Markov chain 
(perturbed by $\epsilon$) over the state space 
$\Omega = ({\mathcal S} \times {\mathcal A})^{K_{max}} \times {\mathcal K} 
\times {\mathcal Q}$ with a positive stationary distribution $\pi_{\epsilon}$.

\begin{proof}
See Appendix \ref{Subsection:Proof of Regular perturbed Markov chain}.
\end{proof}
\end{lemma}

The stationary distribution of the Markov chain $\{ X_{\epsilon}(t) \}$
 characterizes the user associations (the configuration states) 
and the long term average throughput received by the users with the Algorithm~1.
In our work, we seek to characterize the stationary distribution of 
the Markov chain $\{ X_{\epsilon}(t) \}$ especially for small $\epsilon > 0$.
The following definition helps identify the stationary distribution for small 
$\epsilon$ (and the user associations and the average throughput that
occur for a significant fraction of time).
\begin{defn}(Stochastically stable state \cite{young93evolution-of-conventions})
 A state $\omega \in \Omega$ of a regular perturbed Markov chain 
 $\{ X_{\epsilon}(t) \}$ is said to be stochastically stable, if 
 $\lim_{\epsilon \to 0}\pi_{\epsilon}(\omega)>0$.
\end{defn}

We prove optimality by showing that the stochastically stable states of 
$\{ X_{\epsilon}(t) \}$ corresponds to the configuration sequences that
 maximize the network utility. 
 To proceed in that direction, we require an important assumption on the 
 network called interdependence.  

\begin{assum}(Interdependence)
\label{interdependence assum}
For every state $s \in {\mathcal S}$ and for any subset of the users 
$\mathcal{N}^{\prime} \subset \mathcal{N}$ and user association vector 
$\bm{a} = (a_{\mathcal{N}^{\prime}}, a_{ - {\mathcal N}^{\prime}})$, there exists 
a user $j \notin {\mathcal N}^{\prime}$ and a user association vector 
$\bm{a}^{\prime} = (a_{\mathcal{N}^{\prime}}^{\prime},a_{ - {\mathcal N}^{\prime}})$ 
such that $f_j(s,\bm{a}) \neq f_j(s,\bm{a}^{\prime})$.
\end{assum}
\begin{remark}
A key assumption needed for Algorithm~1 to work is the interdependence defined above. We study a completely uncoupled setup where the only feedback to a wireless user on the network configuration is the user's throughput in the slot. The interdependence assumption ensures that changes in user association by any user(s) can be perceived by other users in the network as a change in their user throughput. Algorithm~1 exploits this feature where a discontent user changes associations randomly to effect change in throughput of the other users thereby causing discontent to the other users in the network.
\end{remark}

Further, from \cite{young93evolution-of-conventions}, we know that the 
stochastically stable states of the Markov chain $\{ X_{\epsilon}(t) \}$ must 
necessarily belong to the recurrent classes of Markov chain $\{ X_{0}(t) \}$ 
(the Markov chain obtained by substituting  $\epsilon = 0$ in the transition 
probabilities). The following lemma characterizes the recurrent classes (and states) 
of the Markov chain $\{ X_{0}(t) \}$.

\begin{lemma}
\label{lemma:Recurrent classes}
 The recurrent classes (and states) of the Markov chain $X_0(t)$ are the following:
\begin{enumerate}
\item A state $\omega = (c^{(1)}, \cdots, c^{(K_{max})}, \bm{K}, \bm{q}) \in \Omega$ 
is part of a recurrent class if $q_i = 1$ for all $i \in {\mathcal N}$, 
$s^{(j)}=g(c^{(j-1)})$ and if the association values and the throughput received
repeat with interval $K_i$, for every user $i$.
For example, consider a configuration cycle $(c^{(1)}, \cdots, c^{(K^{\prime})} )$ 
of length $K^{\prime} \leq K_{max}$. Then, $\omega = (c^{(1)}, \cdots, 
c^{(K^{\prime})}, c^{(1)}, \cdots, K^{\prime}, 1)$ is a recurrent state.
The recurrent class to which the state belongs includes states such as
\begin{eqnarray*}
&\hspace{-12mm}((c^{(2)}, \cdots, c^{(K^{\prime})}, c^{(1)}, c^{(2)}, \cdots, K^{\prime}, 1),& \\
&((c^{(3)}, \cdots, c^{(K^{\prime})}, c^{(1)}, c^{(2)}, c^{(3)}, \cdots, K^{\prime}, 1) \cdots .&
 \end{eqnarray*}
 
Let ${\mathcal B} = \{ {\mathcal B}_1, \dots, {\mathcal B}_L \}$ denote the set of all such recurrent classes.
Further, the set ${\mathcal B}$ is non-empty if $K_{\max}\geq |\mathcal{A}\times\mathcal{S}|$. This is because the maximum length of a basic configuration cycle is  $|\mathcal{A}\times\mathcal{S}|$.
\item All states $\omega \in \Omega$ such that $q_i = 0$ for all $i \in {\mathcal N}$ form a single recurrent class. Let us denote this class as ${\mathcal O}$. \hfill \QED
\end{enumerate} 
\begin{proof}
See Appendix \ref{Subsection:Proof of Recurrent classes}
\end{proof}
\end{lemma}

We need the following additional definitions from the theory of regular perturbed 
Markov processes from \cite{young93evolution-of-conventions} to complete the 
discussion on the stochastically stable states of $\{ X_{\epsilon}(t) \}$.

\begin{enumerate}

\item Consider a sequence of state transitions  $\omega_1  \to \dots \to \omega_k$. 
The resistance of the path (sequence of transitions) is defined as the  
sum of the resistances of the one-step transitions in the path,
i.e., $r(\omega_1 , \omega_2)+\dots+r(\omega_{k-1} , \omega_k)$.

\item The resistance from state $\omega_i$ to $\omega_j$ is 
defined as the minimum resistance over all paths from $\omega_i$ to $\omega_j$.

\item The resistance from a recurrent class ${\mathcal B}$ to another 
recurrent class ${\mathcal B}^{\prime}$,
$\rho({\mathcal B},{\mathcal B}^{\prime})$, is defined as the 
minimum resistance from any state $\omega \in {\mathcal B}$ to any state 
$\omega^{\prime} \in {\mathcal B}^{\prime}$.

\item Consider a complete directed graph ${\mathcal G}$ with the recurrent classes
of $\{ X_{0}(t) \}$ as the vertices.  We assign weights to the edges as follows, 
e.g., $\rho({\mathcal B} \rightarrow {\mathcal B}^{\prime})$ is the weight of the 
directed edge from recurrent class $\mathcal{B}$ to $\mathcal{B}^{\prime}$.
 Now, consider a tree rooted at a recurrent class, say ${\mathcal B}_i$,
with a directed path from every other vertex to ${\mathcal B}_i$.
Then, the resistance of the tree is defined as the sum of the weights of the 
edges of the tree.
\item The stochastic potential $\gamma({\mathcal B}_i)$ of a recurrent class 
${\mathcal B}_i$ is defined as the minimum resistance over all trees rooted 
at that recurrent class.
\end{enumerate}

The following lemmas compute the resistance between the recurrent classes of 
$ X_0(t) $ and the stochastic potential of the recurrent classes. 

\begin{lemma}
\label{lemma:resistance bounds}
Consider the recurrent classes ${\mathcal B}_1, \cdots, {\mathcal B}_L$ and ${\mathcal O}$ of
$\{ X_0(t) \}$.
Then,
\begin{enumerate}
	\item $\rho({\mathcal B}_i \rightarrow \mathcal{O}) = z$
	\item Let $(c^{(1)}, \cdots, c^{(K_{max})}, K, 1)$ be a state in ${\mathcal B}_i$. Then,
{\scriptsize	\[ \rho(\mathcal{O} \to {\mathcal B}_i)= \sum_{i=1}^N \left(1-U_i \left(\frac{f_i(c^{(1)}) + \cdots + f_i(c^{(K_i)})}{K_i} \right) \right) \] }
    \item For $j \neq i$, $z \leq \rho(\mathcal{B}_j \to {\mathcal B}_i) < 2z$. \hfill 
\end{enumerate}

\begin{proof}
See Appendix \ref{Subsection:Proof of resistance bounds}.
\end{proof}
\end{lemma} 
\hspace{1 cm} 
\begin{lemma} 
\label{lemma:Stochastic potential of content class}
The stochastic potential of a recurrent class ${\mathcal B}_i$ with state 
$(c^{(1)}, \cdots, c^{(K_{max})}, K, 1)$ is given by
{\scriptsize \[
\gamma({\mathcal B}_i)=z(L-1)+ \sum_{i=1}^N \left(1-U_i \left(\frac{f_i(c^{(1)}) + \cdots + f_i(c^{(K_i)})}{K_i} \right) \right)
\] } \hfill 
\begin{proof}
Follows from Lemma \ref{lemma:resistance bounds} and Lemma $4.3$ in 
\cite{doi:10.1137/110850694}
\end{proof}
\end{lemma}

\begin{lemma} 
\label{lemma:Stochastic potential of discontent class}
The stochastic potential of the recurrent class ${\mathcal O}$ is $L z$ and 
there exists $i$ such that $\rho({\mathcal O}) > \rho({\mathcal B}_i)$. 
\QED
\end{lemma}

The following theorem from \cite{young93evolution-of-conventions} identifies the 
stochastically stable states of the process $\{ X_{\epsilon}(t) \}$ from among 
the recurrent classes of $\{ X_0(t) \}$.
\begin{theorem}\cite{young93evolution-of-conventions}. The stochastically stable 
states of a regular perturbed Markov chain $\{ X_{\epsilon}(t) \}$ are states of the 
recurrent class having minimum stochastic potential. \hfill \QED
\end{theorem}

The above theorem insists that the stochastically stable classes of the Markov chain 
$\{ X_{\epsilon}(t) \}$ are the states where all users are content and that 
minimizes $\gamma({\mathcal B}_i)$, i.e.,
{\scriptsize\[ \sum_{i=1}^N \left(1-U_i \left(\frac{f_i(c^{(1)}) + \cdots + 
f_i(c^{(K_i)})}{K_i} \right) \right). \]}
\noindent
This implies that the stochastically stable states are those that maximize
{\scriptsize\[ \sum_{i=1}^N U_i \left(\frac{f_i(c^{(1)}) + \cdots + f_i(c^{(K_i)})}{K_i} \right). \]}

We note again that any configuration cycle of length $K^{\prime} \leq K_{max}$ 
belongs to the set of recurrent classes of $\{ X_0(t) \}$. 
Hence, the stochastically stable states of the
Markov chain must achieve a sum utility at least as high as these classes (states).
Further, configuration cycles of large lengths permit
almost every convex combination of the basic configuration cycles 
(follows from the irreducibility assumption and the fact that basic configuration 
cycles are of length at most $| {\mathcal S}\times{\mathcal A} |$).
Thus, as $K_{max} \rightarrow \infty$ and as $\epsilon \rightarrow 0$,
the stochastically stable states of Algorithm~1 optimise the formulation 
in (\ref{optimization_problem}).


 \section{Unknown state: Independent state evolution}
\label{Section:Unknown state and Random state transition}
In this section, we assume that the state $s(t)$ is a sequence of independent
random variables drawn with probability mass function (pmf) $\mu(\cdot,\bm{a})$, where $\bm{a}$ 
is the action profile chosen at time~$t$. Thus for a fixed action profile $\bm{a}$,
the state is independent and identically distributed with pmf $\mu(\cdot,\bm{a})$. 
 We also assume that the users do not know the state $s(t)$. 
In this setup, we would like to maximize the following formulation:
\begin{align}
\label{formulation:Unknown State}
\begin{aligned}
\max &\sum_i U_i(\bar{r}_i) \\
\text{ s.t } &\bar{r}_i \leq  \sum_{\bm{a}} p(\bm{a}) E\left(r_i(\bm{a},s)\right),  \\
& \sum_{\bm{a}} p(\bm{a})=1, \; \; p(\bm{a}) \geq 0,
\end{aligned}
\end{align}
where, the expectation is with respect to the distribution $\mu(\cdot,\bm{a})$.
Additionally, in this section, we assume that $U_i$'s are Lipschitz continuous.
\subsection{Utility maximization algorithm}
In this subsection, we shall propose a distributed algorithm to 
maximize the formulation in \eqref{formulation:Unknown State}.
We consider frames of length $L$ slots.
Each user $i$ chooses an action $a_i(l)$ at the 
beginning of every frame $l$ and repeats the same action during the frame.  
Let $\bar{r}_i(l)$ be the average throughput received during frame $l$. 
Users maintain satisfaction variable $q_i(l)$, which is updated at the end of each frame.
We intend to use the algorithm in \cite{7746656} over these frames. 
By choosing frames of suitably large length, the time average throughput 
received over a frame will be close to expected throughput for the chosen 
action profile (expectation over the state $s(t)$). 
We formalize the idea above in the following discussion. 
\par If user $i$ is content at the beginning of frame $l$, it repeats 
the access point it chose $K$ frames earlier with a large probability $1-\epsilon^z$.
If user $i$ is discontent at the beginning of frame $l$, it chooses  
an access points uniformly from the set $\mathcal{A}_i$.
\par If user $i$ was content in the previous frame and repeats 
the same associations chosen $K$ frames before, then 
the player remains content if the difference between the 
average throughput of the current frame and the frame $K$ slots earlier is 
within $\delta$ in magnitude. In other cases, player 
$i$ becomes content with a small probability 
$\epsilon^{1-U_ i\left(\frac{1}{K}\sum\limits_{j=t-K+1}^{t} \bar{r}_i(l) \right)}$, where 
$\bar{r}_i(l)$ is the average throughput received during the frame $l$.      
\begin{algorithm}
\caption{\textbf{: User Association Algorithm}}
\label{Alg3: Unknown state}
\begin{algorithmic}
\Initialize{Fix $z > N$, $K,L \in \mathbb{Z}^+$,  and $\epsilon > 0$. \\ 
For all $i \in {\mathcal N} $, set  $q_i(0) =0.$ }
\end{algorithmic}

\begin{algorithmic}
\State \textbf{Update for user association at frame $l$:}
\If {($q_i(l-1)=1$)}
\State
$a_i(l) =
\begin{cases}
a_i(l-K) & \text{w.p. \ } 1 - \epsilon^z \\
a_i \in A_i & \text{w.p. \ } \frac{ \epsilon^z}{\left|A_i\right|}
\end{cases}$
\Else
\State $a_i(l)=a_i \in A_i$ w.p. $\frac{ 1}{\left|A_i\right|}$
\EndIf
\end{algorithmic}

\begin{algorithmic}
\State \textbf{Update for $q_i(\cdot)$ at time $t$:}
\If { ($q_i(l-1)=1$) and ($a_i(l)=a_i(l-K)$) \\ and $\left( |\bar{r}_i(l) - \bar{r}_i(l-K)|<\delta\right)$}
\State $q_i(l) = 1$
\Else
\State
$q_i(l) =
\begin{cases}
1 & \text{w.p. \ } \epsilon^{1-U_ i\left(\frac{1}{K}\sum\limits_{j=t-K+1}^{t} \bar{r}_i(l) \right)} \\
0 & \text{w.p. \ } 1- \epsilon^{1-U_ i\left(\frac{1}{K}\sum\limits_{j=t-K+1}^{t} \bar{r}_i(l)\right)}
\end{cases}$ \\
\EndIf
\end{algorithmic}
\end{algorithm}

\subsection{Optimality Results}
In this section, we provide sufficient conditions on $L$ and $\delta$ such
that the algorithm maximizes the formulation in \eqref{formulation:Unknown State}
as $\epsilon \to 0$. Before proceeding to the analysis, we modify the Interdependence 
assumption as follows:
\begin{assum}(Interdependence)
\label{iid:interdependence assum}
For any subset of the users $\mathcal{N}^{\prime} \subset \mathcal{N}$ and user association vector 
$\bm{a} = (a_{\mathcal{N}^{\prime}}, a_{ - {\mathcal N}^{\prime}})$, there exists 
a user $j \notin {\mathcal N}^{\prime}$ and a user association vector 
$\bm{a}^{\prime} = (a_{\mathcal{N}^{\prime}}^{\prime},a_{ - {\mathcal N}^{\prime}})$ 
such that $E(r_j(s,\bm{a})) \neq E(r_j(s,\bm{a}^{\prime}))$.
\end{assum}
 \par Choice of $L$ and $\delta$: We choose $L$ and $\delta$ to satisfy the following 
conditions:
\begin{enumerate}
\item $\delta \to 0 \text{ as } \epsilon \to 0$.
\item $L \delta^2 \geq  z \log(1/\epsilon)$
\item $L \to \infty $ as $\epsilon \to 0$. 
\item $L\epsilon^k \to \infty $, for some k. 
\end{enumerate}
{\em One possible choice of $\delta$ and $L$ satisfying the above is $L=1/\epsilon$
and $\delta^2 \geq  z \epsilon \log(1/\epsilon) $.}

\par Let $Z_{\epsilon}(l)=(a_i(l-K+1),\ldots,a_i(l),q_i(l)), i=1,\dots,N)$.
Note that $Z_{\epsilon}(l)$ is a process that changes over frames of length~$L$.  
Now, we have the following lemma,
\begin{lemma}
$Z_{\epsilon}(l)$ is regular perturbed Markov chain on the state space 
$(\mathcal{A}^K \times \mathcal{Q})$.
\QED
\end{lemma}
As a first step in analysing the performance of $Z_{\epsilon}$, we first identify the 
recurrent classes of $Z_{0}(l)$.
\begin{lemma}
The recurrent classes of $Z_{0}(l)$ are as follows:
\begin{enumerate}
\item States where all the users are content. For example, a sequence associations and 
satisfaction variable pair $(\bm{a}^{(1)},\ldots,\bm{a}^{(K)},\bm{1})$ belongs to a recurrent class. 
All cyclic shifts of $(\bm{a}^{(1)},\ldots,\bm{a}^{(K)})$ with all users content also belongs to this class. 
Let $\mathcal{B}_1, \dots, \mathcal{B}_J$ denote the recurrent classes of this type.
\item States where all the users are discontent forms a single recurrent class.
We denote this class by $\mathcal{O}$.  
\end{enumerate} 
\end{lemma}
In the lemma below, we provide bounds on the resistances between the
recurrent classes of $Z_0$.
\begin{lemma}
\label{lemma: Resistance bounds of unknown state}
Let $(\boldsymbol{a}^{(1)},\ldots,\bm{a}^{(K)},1) \in \mathcal{B}_i$. We have the following results,
\begin{enumerate}
\item $\rho(\mathcal{B}_i,\mathcal{O}) = z$. 
Let $\bar{r}_j(l)$ denote the 
average throughput received by user $j$ in frame $l$, when action profile 
$\boldsymbol{a}$ is played. Let  $\hat{r}_j(\boldsymbol{a})$ denote the expected average 
throughput received by user $j$ in a frame i.e., $\hat{r}_j(\boldsymbol{a})=E(\bar{r}_j(l))$.  
A transition from $\mathcal{B}_i$ happens when a user changes its association 
sequence with probability $\epsilon^z$ or when the average throughput 
of a user changes by more than $\delta$ in a frame.
The former happens with resistance at least $z$. 
To calculate the resistance for the latter case, consider,
\begin{align*}
&P\{|\bar{r}_j(l)-\bar{r}_j(l-1)|>\delta\} \\
&= P\{|\bar{r}_j(l)-\hat{r}_j(\boldsymbol{a})+\hat{r}_j(\boldsymbol{a})-\bar{r}_j(l-1)|>\delta\} \\ 
& \leq P\{|\bar{r}_j(l)-\hat{r}_j(\boldsymbol{a})|>\frac{\delta}{2}\} \\
&\overset{(a)}{\le} e^{-\frac{L\delta^2}{2}} \\
&\overset{(b)}{\le} \epsilon^z\\
\end{align*}
Where $(a)$ follows from Hoeffding's lemma and $(b)$ follows from our choice of 
$L \delta^2 \geq  2z \log(\frac{1}{\epsilon})$. 
Thus, $\rho(\mathcal{B}_i,\mathcal{O}) \geq z$. 
By our choice of $\delta$ and interdependence, once a user 
becomes discontent every other player becomes discontent with zero resistance.
Therefore, $\rho(\mathcal{B}_i,\mathcal{O}) = z$. 
\item 
$\text{Let, }\;\overline{\bm{r}}_i = \frac{\sum_{l=1}^K \bar{r}_i(l)}{K}  \text{ and }
\widehat{\bm{r}}_i = \frac{\sum_{l=1}^K \hat{r}_i(\bm{a}^{(l)})}{K}$. Then, we have,
\begin{align*}
\rho(\mathcal{O} \to {\mathcal B}_i)= \sum_{i=1}^N \left(1-U_i \left(\widehat{\bm{r}}_i \right) \right).
\end{align*}
Transition from $\mathcal{O}$ to $\mathcal{B}_i$ would require all the players to 
become content which happens with probability 
$\epsilon^{1-\sum_i U_i\left(\frac{\sum_{l=1}^K \bar{r}_i(l)}{K}\right)}.$  
\par To prove the above, we need to show that, for all $i$,
\begin{align*}
\lim_{\epsilon \to 0}  \epsilon^{U_i\left(\overline{\bm{r}}_i\right)-U_i \left( \widehat{\bm{r}}_i \right) } = 1 
\end{align*} 
By Lipschitz continuity of $U_i$, we have $\forall \delta_1 >0$, with 
probability $1-\epsilon^{-2L\delta_1^2}$, we have, 
\begin{align*}
-P \delta_1 \leq U_i\left(\overline{\bm{r}}_i\right)-U_i \left( \widehat{\bm{r}}_i \right) \leq P \delta_1, 
\end{align*}
where $P$ is assumed to be the Lipschitz constant.
This implies, for all $\delta_1$, we have,   
\begin{align*}
\epsilon^{P\delta_1} ( 1-\epsilon^{-2L\delta_1^2}) 
\leq  \epsilon^{U_i\left(\overline{\bm{r}}_i\right)-U_i \left( \widehat{\bm{r}}_i \right)}, \text{ and }\\
\epsilon^{U_i\left(\overline{\bm{r}}_i\right)-U_i \left( \widehat{\bm{r}}_i \right)} 
\leq \epsilon^{P\delta_1} ( 1-\epsilon^{-2L\delta_1^2}) + \epsilon^{-2L\delta_1^2}
\end{align*}
we have the result by taking the limit along $\delta_1=1/L^{1/4}$.
\item $c \leq \rho(\mathcal{B}_i,\mathcal{B}_j) < 2c $ \\
The proof for the above statement follow from the arguments in \cite{7746656}.
\end{enumerate}
\end{lemma}
\begin{theorem}
\label{theorem:unkown state iid transition}
Under Assumption \ref{iid:interdependence assum} (Interdependence), 
the stochastically stable states of the Markov chain induced by the above algorithm 
are the states which maximize the following formulation, 
\begin{align*}
\max &\sum_i U_i(\bar{r}_i) \\
\text{ s.t } &\bar{r}_i \leq  \sum_a p(a) E\left(r_i(a,s)\right),  \\
& \sum_a p(a)=1, \; \; p(a) \in \{0,\frac{1}{K},\frac{2}{K},\dots,1\}.
\end{align*}
\begin{proof}
Follows from Lemma \ref{lemma: Resistance bounds of unknown state}, 
and Lemmas $4,5$ and Theorem $2$ in \cite{7746656}.
\end{proof}
\end{theorem}
So far we have assumed that the state is unknown to the users. In the following 
sections, we shall assume that the state is known to the users. This assumption
allows us to work with a more general state evolution model (Section \ref{Section:KnownStateControlledMarkov}) and
a significant increase in the rate region (Section \ref{Section:Known state and Random state transition}). 

\section{Known State: Controlled Markov evolution}
\label{Section:KnownStateControlledMarkov}
In this section, we shall assume that the state is known to the users and evolves as a controlled Markov 
process, i.e., 
\begin{align*}
P(s(t+1)|s(j),\bm{a}(j), 0\leq j \leq t ) = P(s(t+1)|s(t),\bm{a}(t)) 
\end{align*} 
We say that the control $\bm{a}(t)$ is stationary, if it satisfies 
\begin{align*}
\bm{a}(t)=\bm{h}(s(t)) = (h_1(s(t)),\ldots,h_N(s(t))),
\end{align*}
where $\bm{h}$ is a deterministic function from $\mathcal{S}$ to $\mathcal{A}$.
We assume that the for any stationary control $\bm{h}$, the controlled Markov process 
$S(t,\bm{h})$ is ergodic with stationary distribution $\mu(\cdot,\bm{h})$. 
Further, for a given control $\bm{h}$, the expected stationary pay-off is given by,
\begin{align}
\label{Controlled Markov:Expected payoff}
r_i(\bm{h}) = \sum_{s \in \mathcal{S}} \mu(s,\bm{h}(s)) r_i(s,\bm{h}(s))
\end{align}
Denote by $\mathcal{H}$ the set of stationary controls. Since the set of action profiles and states are finite, 
the set $\mathcal{H}$ is finite as well. Our objective here is to time share between functions $\bm{h}$ such that the  
sum utility is maximized. Formally,
\begin{align}
\begin{aligned}
\label{formulation:Controlled Markov process}
\max &\sum_i U_i(\bar{r}_i) \\
&\text{s.t.  } \bar{r}_i = \sum_{\bm{h} \in \mathcal{H}} p(\bm{h}) r_i(\bm{h}) \\
& \sum_{\bm{h}} p(\bm{h}) = 1, \; \; p(\bm{h}) \in \left\{0,\frac{1}{K},\cdots,1 \right\}
\end{aligned}
\end{align}
Note the similarity between the above formulation and \eqref{formulation:Unknown State}. In the formulation
above, the expected pay-off in \eqref{Controlled Markov:Expected payoff} is with respect to the stationary distribution
$\mu(s,\bm{h})$ of the controlled Markov chain, whereas in \eqref{formulation:Unknown State} we assumed that the expectation 
is with respect to an $iid$ random variable with distribution $\mu$. Hence, to solve the above formulation, we run 
Algorithm~\ref{Alg3: Unkown state}, where node $i$ chooses stationary control $h_i$    
To establish an estimate on the frame size $L$ and $\delta$, we need the following assumptions on the controlled Markov chain.
\begin{assum}
For each $\bm{h} \in \mathcal{H}$, the controlled Markov chain $S(t,\bm{h})$ has the following property,
\begin{align*}
 \lambda_2(\bm{h}) <  \lambda_{\min},
\end{align*}
where $\lambda_2(\bm{h})$ is the second largest eigenvalue modulus of the controlled Markov chain $S(t,\bm{h})$. 
\end{assum}
Now we have the results of Theorem \ref{theorem:unkown state iid transition} holds for the formulation \eqref{formulation:Controlled Markov process} with 
the following choices of $L$ and $\delta$ with appropriate interdependence assumption (i.e., with $\bm{a}$ replaced by $\bm{h}$ and 
$E(r_i(\bm{a},s))$ replaced by $r_i(\bm{h})$ in Assumption \ref{iid:interdependence assum}).
\begin{enumerate}
\item $\delta \to 0 \text{ as } \epsilon \to 0$.
\item $L \delta^2 \geq  \frac{z(1-\lambda_{\min})}{1+\lambda_{\min}} \log(1/\epsilon)$
\item $L \to \infty $ as $\epsilon \to 0$. 
\item $L\epsilon^k \to \infty $, for some k. 
\end{enumerate}
The proof follows by replacing the Hoeffding inequality for $iid$ random variables with the inequality for Markov 
chains (See Theorem $2.3$ in \cite{fan2018hoeffding}). 

 \section{Known state: Ergodic state evolution}
\label{Section:Known state and Random state transition}
In this section, we shall assume that the state $s(t)$ evolves as an ergodic 
random process taking values in a finite set $\mathcal{S}$ with time average 
probabilities $\mu(\cdot)$.
We assume that the users know the state $s(t)$ prior to choosing their
 associations at time~$t$. 
 In this setup, we aim to maximise the following formulation:
\begin{align*}
\max &\sum_i U_i(\bar{r}_i) \\
\text{ s.t } &\bar{r}_i \leq \sum_{s\in \mathcal{S}} \mu(s) \sum_a p(a,s) r_i(a,s),  \\
& \sum_a p(a,s)=1, \; \; p(a,s) \geq 0, \; \forall s \in S.
\end{align*}
In the following subsections, we describe the proposed algorithm
 and discuss optimality results.
\subsection{Utility maximization algorithm}
We now propose a completely uncoupled utility 
maximization algorithm, assuming that users know the state prior to 
choosing access points.
 Each user has a binary satisfaction variable $q_i(t)$.
The purpose of $q_i(t)$ is similar to the algorithm with 
deterministic state transition in Section \ref{Subsection:Deterministic Algorithm}. 
A user chooses an access point based on the current and prior state,
 history of the access points chosen by the user and its satisfaction variable $q_i$.  
 Let the history of system state, the access points chosen and throughput received 
 by user~$i$ until time $t$  be $\{(s(l),a_i(l),r_i(l)),\; l=1, \dots ,t-1\}$. 
 For each state $s$, we require the users to keep track of associations 
 and throughput received during  the last $K$ occurrences of state $s$. 
 We denote by $(\hat{a}_i(K,s),\hat{r}_i(K,s))$, the access point chosen and 
 throughput received by user $i$ the previous time when state $s$ occurred. 
 Let $\{(\hat{a}_i(j,s),\hat{r}_i(j,s)), j=1,\dots,K\}$ denote access points chosen
  and throughput received by user $i$ during the $K$ recent time slots when 
  state $s$ occurred. We require each user to keep track of the history
   $\{(\hat{a}_i(j,s),\hat{r}_i(j,s)), j=1,\dots,K\}$ for every state $s \in \mathcal{S}$. 
  If state $s$ occurred for less than $K$ times, we set by default,   
  $\hat{a}_i(j,s)= a_0 \in \mathcal{A}_i$, and $\hat{r}_i(j,s)=0$, for all $j$, 
  where state $s$ has occurred for less than $K-j+1$ times. We 
  also require each user to keep track of the number of times state 
  $s$ has occurred and denote it by $t_s$.
  Then, $t_s/t$ denotes the fraction of time state $s$ has occurred. 
  \par Recall that, we have assumed that every user knows the state
  before choosing the access point to associate with.  
  If user $i$ was content in slot $t-1$ and  
  the current state is $s(t)$, then user $i$ chooses the access point $\hat{a}_i(1,s(t))$
  with a large probability ($1-\epsilon^z$). Here, $\hat{a}_i(1,s(t))$ is the 
  access point chosen by user $i$ the $K$th last time state $s(t)$ occurred.  
  With a small probability $\epsilon^z$, user $i$ chooses any other access point 
  uniformly at random. If user $i$ was discontent in slot $t-1$, 
  then it chooses an access point uniformly at random from $\mathcal{A}_i$  
  independent of the state $s(t)$. 
  \par User $i$ updates its satisfaction variable $q_i(t)$ based on the
   fraction of time each state has occurred $(\{t_s/t\}\; s \in S)$, the current state 
   $s(t)$, and its prior satisfaction variable $q_i(t-1)$,  
   history $(\hat{a}_i(K,s),\hat{r}_i(K,s))$, current association $a_i(t)$ and 
   throughput $r_i(t)$. If player $i$ was content in slot~$t-1$, and chose 
   the action $\hat{a}_i(1)$ and received the payoff $\hat{r}_i(1)$ in slot~$t$,   
   then it remains content ($q_i(t)=1$) with probability $1$. 
   In other cases, player $i$ becomes content $(q_i(t)=1)$ 
   with a small probability $\epsilon^{1-U_i(\bar{r}_i)}$. 
   Here, $\bar{r}_i$ is given as follows. Let $\bar{r}_i(s)$ denote the average payoff
   received by player $i$ over the previous $K$ slots when state $s$ occurred
   i.e., $\bar{r}_i(s)= 1/K  \sum_{j=1}^{K} \hat{r}_i(j,s) $. Now $\bar{r}_i$ is
   the weighted average of $\bar{r}_i(s)$ weighted by the fraction of time 
   state $s$ has occurred i.e.,
   \begin{align*}
   \bar{r}_i = \sum\limits_{s\neq s(t)} \frac{t_{s}}{t} \bar{r}_i(s) + \frac{t_{s(t)}}{t}   
   \frac{1}{K}\left(\sum\limits_{j=2}^{K} \hat{r}_i(j,s(t))+r_i(t)\right).
   \end{align*}
   Finally, $(\hat{a}_i(j,s(t)),\hat{r}_i(j,s(t)))$ is updated with the recent action 
   and payoff. 
\begin{algorithm}
\caption{\textbf{: User Association Algorithm}}
\label{Alg2: Known state}
\begin{algorithmic}
\Initialize{Fix $z > N$, $K \in \mathbb{Z}^+$ and $\epsilon > 0$. \\ 
 For all $i \in {\mathcal N} $, $j = 1, \dots, K$, and  $s \in S$, \\ set $\hat{a}_i(j,s)= a_0 \in \mathcal{A}_i$, $\hat{r}_i(j,s)=0$, $q_i(0) =0.$ }
\end{algorithmic}

\begin{algorithmic}
\State \textbf{Update for State at time $t$:}
\State $t_{s(t)}= t_{s(t)}+1$
\end{algorithmic}
\begin{algorithmic}
\State \textbf{Update for user association at time $t$:}
\If {($q_i(t-1)=1$)}
\State
$a_i(t) =
\begin{cases}
\hat{a}_i(1,s(t)) & \text{w.p. \ } 1 - \epsilon^z \\
a_i \in A_i & \text{w.p. \ } \frac{ \epsilon^z}{\left|A_i\right|}
\end{cases}$
\Else
\State $a_{i}(t)= a_i \in A_i$ w.p. $\frac{ 1}{\left|A_i\right|}$
\EndIf
\end{algorithmic}

\begin{algorithmic}
\State \textbf{Update for $q_i(\cdot)$ at time $t$:}
\If { ($q_i(t-1)=1$) and ($a_i(t)=\hat{a}_i(1,s(t))$) \\ and $\left( r_i(t) = \hat{r}_i(1,s(t))\right)$}
\State $q_i(t) = 1$
\Else
\State
$q_i(t) =
\begin{cases}
1 & \text{w.p. \ } \epsilon^{1-U_ i\left(\bar{r}_i(t) \right)} \\
0 & \text{w.p. \ } 1- \epsilon^{1-U_ i\left(\bar{r}_i(t)\right)}
\end{cases}$ \\
where, \\
   $\bar{r}_i = \sum\limits_{s\neq s(t)} \frac{t_{s}}{t} \bar{r}_i(s) + \frac{t_{s(t)}}{t}\frac{1}{K}\left(\sum\limits_{j=2}^{K} \hat{r}_i(j,s(t))+r_i(t)\right)$, \\
   $\bar{r}_i(s)= \frac{1}{K}  \sum\limits_{j=1}^{K} \hat{r}_i(j,s)$.
\EndIf
\end{algorithmic}

\begin{algorithmic}
\State \textbf{Update for $\hat{a}_i(\cdot,\cdot)$ and $\hat{r}_i(\cdot,\cdot)$ at time $t$:}
\State For $j=1, \dots, K-1$, 
 set  $\hat{a}_i(j,s(t)) = \hat{a}_i(j+1,s(t))$, \\ 
 and $\hat{r}_i(j,s(t)) = \hat{r}_i(j+1,s(t))$.
\State Set $\hat{a}_i(K,s(t))=a_i(t)$ and $\hat{r}_i(K,s(t))=r_i(t)$
\end{algorithmic}
\end{algorithm}

 \subsection{Optimality Results}
In this subsection, we will study the stationary performance of Algorithm 2 as 
$\epsilon \to 0$. Let $Y_{\epsilon}(t)=(\{(\hat{a}_i(j,s),q_i(t)),\;  j=1\dots K, \; s\in 
\mathcal{S},\; i \in \mathcal{N}\})$. 
 First, we will show in the lemma below that the algorithm induces a Markov chain. 
\begin{lemma}
\label{lemma:Alg2MarkovChain}
$Y_{\epsilon}(t)$ induces a time non-homogeneous Markov chain on the state space 
$\mathcal{A}^{K \times|\mathcal{S}|}\times\mathcal{Q}$. 
\begin{proof}
See Appendix \ref{Subsection:Proof of Alg2MarkovChain}
\end{proof}
\end{lemma}
Let $P_{\epsilon}(t)$ denote the transition probability matrix of $Y_{\epsilon}(t)$. 
Also, let $\hat{P}_{\epsilon}$ denote the transition probability matrix of 
algorithm 2 with $t_s/t$ replaced by its ensemble average $\mu(s)$. 

In the next lemma we show that the Markov chain is strongly ergodic.
\begin{defn}
A non-homogeneous Markov chain with with transition probability matrix
$P(t)$ is strongly ergodic if there exists a probability distribution $\pi$, such that, 
for all $m\geq0$, we have, 
\begin{align*}
\lim_{k \to \infty} \sup_{\mu} d_V(\mu^T P(m,k), \pi) =0,
\end{align*}
where, $P(m,k) = \prod_{j=m}^{k-1} P(j)$ and $d_V(\cdot,\cdot)$ is the 
total variation distance.
\end{defn}
\begin{lemma}
The Markov chain $Y_{\epsilon}(t)$ is strongly ergodic. 
\begin{proof}
By ergodicity of $s(t)$, we have, $\lim_{t \to \infty} t_s/t = \mu_s$.  
Also with continuity of $U_i's$, we have,
\begin{align*}
\lim_{t \to \infty}|P_{\epsilon}(t) - \hat{P}_{\epsilon}| =0
\end{align*}
Note that $\hat{P}_{\epsilon}$ is an ergodic transition probability matrix.
Thus, by Theorem $V.4.5$ in \cite{isaacson1976markov}, the Markov chain 
$Y_{\epsilon}(t)$ is strongly ergodic.
\end{proof}
\end{lemma}
The theorem below characterizes the stationary performance of the 
Markov chain $Y_{\epsilon}$ as $\epsilon \to 0$.
\begin{theorem}
Under Assumption \ref{interdependence assum} (Interdependence), 
the stochastically stable states of the Markov chain $Y_{\epsilon}(t)$ 
maximizes the following formulation:
\begin{align*}
\begin{aligned}
& \max \sum_i U_i(\bar{r}_i) \\
\text{s.t } &\bar{r}_i \leq \sum_{s\in \mathcal{S}} \mu(s) \sum_a p(a,s) r_i(a,s),  \\
& \sum_a p(a,s)=1, \; \; p(a,s) \in \left\{0,\frac{1}{K},\frac{2}{K},\ldots,1\right\}, \; \forall s \in S.
\end{aligned}
\end{align*}
\begin{proof}
The stochastically stable states of $Y_{\epsilon}(t)$ is the stochastically stable states 
of $\hat{P}_{\epsilon}$.
The proof follows similar to Theorem 2 in \cite{7746656} for $\hat{P}_{\epsilon}$. 
\end{proof}
\end{theorem}

\section{Numerical Examples} 
In this section, we shall present numerical simulation of our proposed algorithms
in the context of user association in IEEE 802.11ac WiFi network.
The simulations are performed using a ns3/c++ simulator. We assume that 
access points independently choose their channel and their channel choice is modeled as 
the state of the network. We consider an IEEE 802.11ac WiFi network with 
three access points and five users. The access points are placed at the vertices of an 
equilateral triangle of length $25$ meters. We assume that, two orthogonal $20$ MHz channels
are available and in each time slot, the access points can operate in one of them. 
We consider three states, where each state corresponds to allocating an orthogonal 
channel to an access point and the other two access points share a common channel. 
For example, state $1$ corresponds to allocating an orthogonal channel to access point $1$, 
whereas access points $2$ and $3$ share a common channel. In each time slot, the objective 
of our algorithm is to choose user association decisions that maximizes the sum utility of 
the users. In this example, we shall consider the utility
$\log(\delta+\bar{r}_i)$ (for a small $\delta>0$). The $\log$ utility is shown to achieve 
proportional fairness in \cite{Kelly1998} and we use $\log(\delta + \bar{r}_i)$ to keep 
the utility function bounded.
\par
\begin{figure}[!ht]
\begin{center}
\begin{tikzpicture}
\pgfplotstableread{Trunc_10_9/Alg1/K2/100Trunc_0_05_K2_utility_10_9.txt} \exa
\pgfplotstableread{Trunc_10_9/Alg1/K2/100Trunc_0_1_K2_utility_10_9.txt} \exb
\pgfplotstableread{Trunc_10_9/Alg1/K2/100Trunc_0_15_K2_utility_10_9.txt} \exc
\pgfplotstableread{Trunc_10_9/Alg1/K2/100Trunc_0_2_K2_utility_10_9.txt} \exd
\pgfplotstableread{Trunc_10_9/Alg1/K2/100Trunc_0_25_K2_utility_10_9.txt} \exe
\pgfplotstableread{Trunc_10_9/Alg1/K2/100Trunc_0_3_K2_utility_10_9.txt} \exf
\pgfplotstableread{Trunc_10_9/Alg1/K2/100Trunc_0_35_K2_utility_10_9.txt} \exg
\pgfplotstableread{Trunc_10_9/Alg1/ExactgradAlg1_9.txt} \exacta
\begin{axis}[
name=plot1,
ymin=0.48, ymax=0.58,
grid=major,
grid style={dashed, draw=gray!70},
xlabel={Number of slots},
ylabel={Sum Utility, $\sum_i \log(\delta+\bar{r}_i)$},
legend style={at={(0.65,0.5)}, anchor=north,legend columns=1}
]
\addplot+[mark=none,blue,very thick] table[x=Slots,y=SumUtility]\exa;
\addplot+[mark=none,red,dotted,very thick] table[x=Slots,y=SumUtility]\exb;
\addplot+[mark=none,color={rgb,255:red,32;green,127;blue,43},dashed,very thick] table[x=Slots,y=SumUtility]\exd;
\addplot+[mark=none,violet,densely dotted,very thick] table[x=Slots,y=SumUtility]\exf;
\addplot+[mark=none,black,loosely dashed,very thick] table[x=Slots,y=SumUtility]\exacta;
\legend{$\epsilon = 0.05$,$\epsilon = 0.1$ ,$\epsilon = 0.2$,$\epsilon = 0.3$,Gradient Algorithm}
\end{axis}
\end{tikzpicture}
\caption{Sum Utility of the users obtained by Algorithm \ref{Alg1: Deterministic case}, for an IEEE 802.11ac WiFi 
network with 5 users and 3 Access points. The state corresponds to channel allocated to the access points
and the state transition is deterministic. The performance of a centralized sub gradient algorithm 
is shown for reference.} 

\label{Plot:User Association Deterministic}
\end{center}
\end{figure}
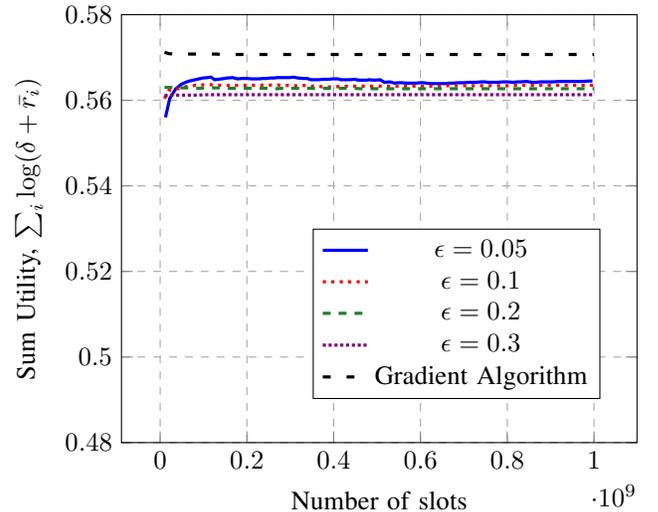

 For the deterministic state transition case, we assume that orthogonal channel is 
allocated to the access point with the maximum number of users. We also assume that 
ties are resolved in a deterministic manner. We run Algorithm $\ref{Alg1: Deterministic case}$
for different values of $\epsilon$ with $K_{\max} = 2$.
In Figure \ref{Plot:User Association Deterministic}, we plot the sum utility of users for 
$\epsilon = 0.05$, $0.1$, $0.2$, and $0.3$. We also plot the performance of a centralized subgradient 
algorithm for reference.    

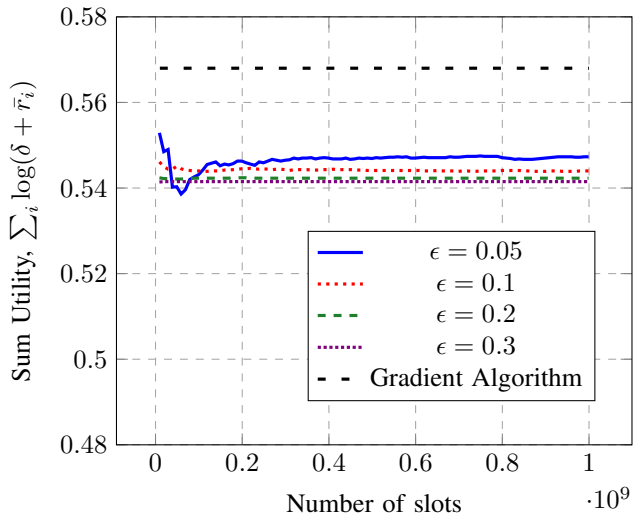
\begin{figure}[!ht]
\begin{center}
\begin{tikzpicture}
\pgfplotstableread{Trunc_10_9/Alg3/K2/Fram2000/100Trunc_0_05_K2_utility_10_9.txt} \exo
\pgfplotstableread{Trunc_10_9/Alg3/K2/Fram2000/100Trunc_0_1_K2_utility_10_9.txt} \exp
\pgfplotstableread{Trunc_10_9/Alg3/K2/Fram2000/100Trunc_0_15_K2_utility_10_9.txt} \exq
\pgfplotstableread{Trunc_10_9/Alg3/K2/Fram2000/100Trunc_0_2_K2_utility_10_9.txt} \exr
\pgfplotstableread{Trunc_10_9/Alg3/K2/Fram2000/100Trunc_0_25_K2_utility_10_9.txt} \exs
\pgfplotstableread{Trunc_10_9/Alg3/K2/Fram2000/100Trunc_0_3_K2_utility_10_9.txt} \ext
\pgfplotstableread{Trunc_10_9/Alg3/K2/Fram2000/100Trunc_0_35_K2_utility_10_9.txt} \exu
\pgfplotstableread{Trunc_10_9/Alg3/ExactGradient/ExactgradAlg3_9.txt} \exactc
\begin{axis}[
name=plot1,
ymin=0.48, ymax=0.58,
grid=major,
grid style={dashed, draw=gray!70},
xlabel={Number of slots},
ylabel={Sum Utility, $\sum_i \log(\delta+\bar{r}_i)$},
legend style={at={(0.65,0.5)}, anchor=north,legend columns=1}
]
\addplot+[mark=none,blue,very thick] table[x=Slots,y=SumUtility]\exo;
\addplot+[mark=none,red,dotted,very thick] table[x=Slots,y=SumUtility]\exp;
\addplot+[mark=none,color={rgb,255:red,32;green,127;blue,43},dashed,very thick] table[x=Slots,y=SumUtility]\exr;
\addplot+[mark=none,violet,densely dotted,very thick] table[x=Slots,y=SumUtility]\ext;
\addplot+[mark=none,black,loosely dashed,very thick] table[x=Slots,y=SumUtility]\exactc;
\legend{$\epsilon = 0.05$,$\epsilon = 0.1$ ,$\epsilon = 0.2$,$\epsilon = 0.3$, Gradient Algorithm}
\end{axis}
\end{tikzpicture}
\caption{Sum Utility of the users obtained by Algorithm \ref{Alg3: Unknown state}, for an IEEE 802.11ac WiFi 
network with 5 users and 3 Access points. The state corresponds to channel allocated to the access points, 
the state transition is $iid$ and the state is unknown to the users. The performance of a centralized subgradient algorithm 
is shown for reference.} 

\label{Plot:User Association Unknown state}
\end{center}
\end{figure}

For the other two cases, we assume that channels are allocated independent of 
user associations. In a fair channel allocation, each access point gets an equal time 
share of the orthogonal channel. 
In every time slot, choosing a state uniformly at random correspond to equal 
time sharing of the orthogonal channel between access points. Thus, we assume that, 
the state evolution is $iid$ and uniformly distributed. 
In the second example, we assume that
channel allocation is unknown to the users prior to association. 
We run Algorithm~$\ref{Alg3: Unknown state}$ with $K = 2$, $L = 4000$ and 
$\delta = 0.05 $. We plot the sum utility for $\epsilon = 0.05$, $0.1$, $0.2$, and
$0.3$ in Figure \ref{Plot:User Association Unknown state}. We also plot the performance of a centralized subgradient algorithm for reference.

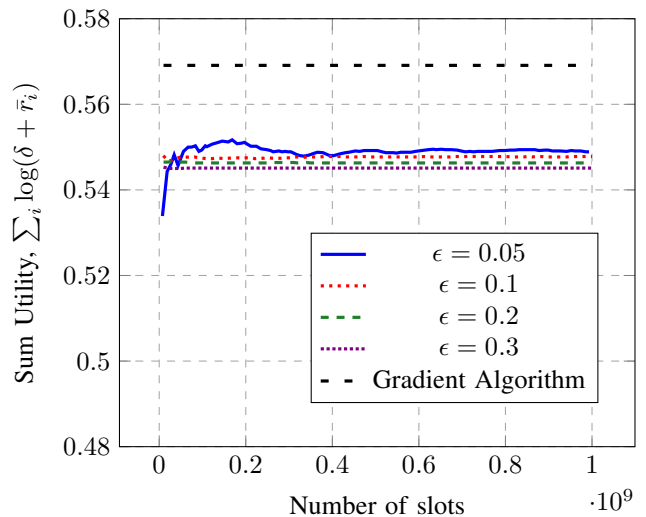
\begin{figure}[!ht]
\begin{center}
\begin{tikzpicture}
\pgfplotstableread{Trunc_10_9/Alg2/K2/100Trunc_0_05_K2_utility_10_9.txt} \exh
\pgfplotstableread{Trunc_10_9/Alg2/K2/100Trunc_0_1_K2_utility_10_9.txt} \exi
\pgfplotstableread{Trunc_10_9/Alg2/K2/100Trunc_0_15_K2_utility_10_9.txt} \exj
\pgfplotstableread{Trunc_10_9/Alg2/K2/100Trunc_0_2_K2_utility_10_9.txt} \exk
\pgfplotstableread{Trunc_10_9/Alg2/K2/100Trunc_0_25_K2_utility_10_9.txt} \exl
\pgfplotstableread{Trunc_10_9/Alg2/K2/100Trunc_0_3_K2_utility_10_9.txt} \exm
\pgfplotstableread{Trunc_10_9/Alg2/K2/100Trunc_0_35_K2_utility_10_9.txt} \exn
\pgfplotstableread{Trunc_10_9/Alg2/ExactgradAlg2_9.txt} \exactb
\begin{axis}[
name=plot1,
ymin=0.48, ymax=0.58,
grid=major,
grid style={dashed, draw=gray!70},
xlabel={Number of slots},
ylabel={Sum Utility, $\sum_i \log(\delta+\bar{r}_i)$},
legend style={at={(0.65,0.5)}, anchor=north,legend columns=1}
]
\addplot+[mark=none,blue,very thick] table[x=Slots,y=SumUtility]\exh;
\addplot+[mark=none,red,dotted,very thick] table[x=Slots,y=SumUtility]\exi;
\addplot+[mark=none,color={rgb,255:red,32;green,127;blue,43},dashed,very thick] table[x=Slots,y=SumUtility]\exk;
\addplot+[mark=none,violet,densely dotted,very thick] table[x=Slots,y=SumUtility]\exm;
\addplot+[mark=none,black,loosely dashed,very thick] table[x=Slots,y=SumUtility]\exactb;
\legend{$\epsilon = 0.05$,$\epsilon = 0.1$ ,$\epsilon = 0.2$,$\epsilon = 0.3$, Gradient Algorithm}
\end{axis}
\end{tikzpicture}
\caption{Sum Utility of the users obtained by Algorithm \ref{Alg2: Known state}, for an IEEE 802.11ac WiFi 
network with 5 users and 3 Access points. The state corresponds to channel allocated to the access points, 
the state transition is $iid$ and is known to the users. The performance of a centralized subgradient algorithm 
is shown for reference.} 

\label{Plot:User Association known state}
\end{center}
\end{figure}

In the third case, 
we assume that the channel allocation is known to the users prior to association.
We run Algorithm \ref{Alg2: Known state} for different values of $\epsilon$ with 
$K = 2$. We plot the sum utility for $\epsilon = 0.05$, $0.1$, $0.2$, and $0.3$ in
Figure \ref{Plot:User Association known state}. We also plot the sum utility
obtained by a centralized subgradient Algorithm for reference.
 
\section{Conclusion}
In this work, we present completely uncoupled 
utility maximisation algorithms for a state based network model. We have considered 
four cases based on the knowledge of the state and its evolution.
We further presented the performance of these algorithms for user 
association, where the state corresponds to channels in which 
the access points operate.
\par In our earlier work \cite{8629359},
we have presented a completely uncoupled subgradient algorithm for 
maximizing concave utilities. We conclude by noting that, with modifications 
as considered in this paper, we could extend the subgradient algorithm in \cite{8629359}
to a state based model as well.

 \section{Appendix}
\subsection{Proof of Lemma \ref{Regular perturbed Markov chain}}
\label{Subsection:Proof of Regular perturbed Markov chain}
We know that $\{ X_{\epsilon}(t) \}$ is a discrete time, finite state 
space random process. At time $t+1$ and for any user $i$, the transition
 probabilities for $a_i(t+1)$ are a function only of $q_i(t), K_i(t)$ and 
 $a_i(t+1-K_i(t))$ (i.e., the current state $X_{\epsilon}(t)$).
And, the transition probabilities for $s(t+1)$ are a function only of 
$s(t)$ and $a(t+1)$ (and hence a function of the current state $X_{\epsilon}(t)$).
Also, the transition probabilities for $q_i(t+1)$ (and $K_i(t+1)$) 
is a function only of $q_i(t)$, $K_i(t)$ and the configuration
states $c(t+2-K_{max}), \cdots c(t+1)$ (the throughputs are a deterministic function
of the user association vectors and the system states).
Hence, we conclude that the
transition probabilities of $\{ X_{\epsilon}(t) \}$ are independent 
of the past, given $X_{\epsilon}(t)$. Thus, $\{ X_{\epsilon}(t) \}$ is a Markov chain.
Also, for any $\epsilon > 0$, $\{ X_{\epsilon}(t) \}$ is an irreducible 
and aperiodic random process (follows from the irreducibility assumption 
of the system state and the transition probabilities in Algorithm~\ref{Alg1: Deterministic case}).
Thus, for any $\epsilon >0$, $\{ X_{\epsilon}(t) \}$ is an ergodic Markov process. 
Let $\pi_{\epsilon}$ denote the unique (and positive) stationary 
distribution of $\{ X_{\epsilon}(t) \}$.

\par From the state transition probabilities listed in Algorithm~\ref{Alg1: Deterministic case}, 
we clearly see that conditions 2) and 3) are satisfied as well.
Hence, $\{ X_{\epsilon}(t) \}$ is a regular perturbed Markov 
chain (perturbed by $\epsilon$).
\QED
\subsection{Proof of Lemma \ref{lemma:Recurrent classes}}
\label{Subsection:Proof of Recurrent classes}
When $\epsilon =0$, a content user repeats the action it chose $K_i$ slots before.   
Also, if a content user receives the payoff that it received $K_i$ slots before, 
then it remains content. Thus, any state $(c^{(1)},\dots,c^{(K_{\max})},K,\vec{1})$ 
where, all the users are content and the association values and throughput 
received repeat with interval $K_i$ ( for every user $i$ ) is a recurrent state in $X_0$.
\par When all the users are discontent, users choose actions uniformly at random.
Due to assumption \ref{assum:Irreducibility} (Irreducibility), there is a positive 
probability of reaching all possible configurations. Hence, the set $\mathcal{O}$ is a 
recurrent class.
\par Consider any state with at least one discontent user. For a content user to 
remain content, the payoff it receives should repeat every $K_i$ slots. 
However, by assumption \ref{interdependence assum} (Interdependence), 
the discontent users could choose actions such that a content user(s) experiences 
a change in payoff forcing the content user(s) to become discontent. 
Extending this argument, all the users will become discontent with a 
positive probability. Thus, a state with some content and rest discontent users is not a 
recurrent class of $X_{\epsilon}$.    
 \QED
 \subsection{Proof of Lemma \ref{lemma:resistance bounds}}
 \label{Subsection:Proof of resistance bounds}
 \begin{enumerate}
 \item A transition from $\mathcal{B}_i$ to $\mathcal{O}$ involves at least one user 
 to change its action and hence become discontent. This happens with resistance $z$.
 Once a user is discontent,  every other user could become discontent with zero 
 resistance (due to interdependence). Thus $\rho(\mathcal{B}_i,\mathcal{O})=z$.
 \item A transition from $\mathcal{O}$ to $\mathcal{B}_i$ involves all the users
 becoming content. User $i$ becomes content with resistance 
 $(1-U_i\left(\frac{f_i(c^{(1),\dots,c^{(K_i)}})}{K_i}\right))$.
 \item A transition from $\mathcal{B}_j$ at least one user becoming discontent with 
 resistance $z$. The upper bound follows from: 
 $\rho(\mathcal{B}_j ,\mathcal{B}_i) \leq \rho(\mathcal{B}_j ,\mathcal{O}) 
 +\rho(\mathcal{O} ,\mathcal{B}_i)$. 
 \end{enumerate}
 \QED
 \subsection{Proof of Lemma \ref{lemma:Alg2MarkovChain}}
 \label{Subsection:Proof of Alg2MarkovChain}
The action chosen at time t, $a(t)$ depends on $q(t-1)$ and $\hat{a}$ at time $t-1$. 
The update of $\hat{a}$ at time $t$ depends only on $\hat{a}$ at time $t-1$ and 
the action $a(t)$ chosen at time $t$.
Also, the satisfaction variable $q(t)$ depends on $q(t-1)$, $a(t)$, fraction of time 
each state occured $t_s/t$ and $\hat{a}$. 
(Note that $r(t)=f(a(t))$ and $\hat{r}(t)=f(\hat{a}(t))$).
Thus $Y_{\epsilon}(t)$ is a Markov chain. The Markov chain is time 
non homogeneous due to the explicit time dependence in $t_s/t$.


\bibliographystyle{IEEEtran}
\bibliography{IEEEabrv,welfare-maximization}

\begin{thebibliography}{10}
\providecommand{\url}[1]{#1}
\csname url@samestyle\endcsname
\providecommand{\newblock}{\relax}
\providecommand{\bibinfo}[2]{#2}
\providecommand{\BIBentrySTDinterwordspacing}{\spaceskip=0pt\relax}
\providecommand{\BIBentryALTinterwordstretchfactor}{4}
\providecommand{\BIBentryALTinterwordspacing}{\spaceskip=\fontdimen2\font plus
\BIBentryALTinterwordstretchfactor\fontdimen3\font minus
  \fontdimen4\font\relax}
\providecommand{\BIBforeignlanguage}[2]{{%
\expandafter\ifx\csname l@#1\endcsname\relax
\typeout{** WARNING: IEEEtran.bst: No hyphenation pattern has been}%
\typeout{** loaded for the language `#1'. Using the pattern for}%
\typeout{** the default language instead.}%
\else
\language=\csname l@#1\endcsname
\fi
#2}}
\providecommand{\BIBdecl}{\relax}
\BIBdecl

\bibitem{7925713}
S.~{Ramakrishnan}, V.~{Ramaiyan}, and K.~P. {Naveen}, ``A distributed user
  association algorithm for state dependent wireless networks,'' in \emph{2017
  IEEE Wireless Communications and Networking Conference (WCNC)}, March 2017,
  pp. 1--6.

\bibitem{marden-etal12pareto-optimality}
J.~R. Marden, H.~P. Young, and L.~Y. Pao, ``{Achieving Pareto Optimality
  Through Distributed Learning},'' in \emph{51st IEEE Conference on Decision
  and Control (CDC)}, Dec 2012, pp. 7419--7424.

\bibitem{1374962}
M.~J. Neely, E.~Modiano, and C.~E. Rohrs, ``Dynamic power allocation and
  routing for time-varying wireless networks,'' \emph{IEEE Journal on Selected
  Areas in Communications}, vol.~23, no.~1, pp. 89--103, Jan 2005.

\bibitem{Kelly1998}
F.~P. Kelly, A.~K. Maulloo, and D.~K.~H. Tan, ``{Rate control for communication
  networks: shadow prices, proportional fairness and stability},''
  \emph{Journal of the Operational Research Society}, vol.~49, no.~3, pp.
  237--252, Mar 1998.

\bibitem{1310314}
H.~J. Kushner and P.~A. Whiting, ``{Convergence of proportional-fair sharing
  algorithms under general conditions},'' \emph{IEEE Trans. Wireless Commun.},
  vol.~3, no.~4, pp. 1250--1259, 2004.

\bibitem{DBLP:journals/ton/JiangW10}
L.~Jiang and J.~C. Walrand, ``{A Distributed {CSMA} Algorithm for Throughput
  and Utility Maximization in Wireless Networks},'' \emph{{IEEE/ACM} Trans.
  Netw.}, vol.~18, no.~3, pp. 960--972, 2010.

\bibitem{5625654}
L.~Jiang, D.~Shah, J.~Shin, and J.~Walrand, ``Distributed random access
  algorithm: Scheduling and congestion control,'' \emph{IEEE Trans. Inf.
  Theory}, vol.~56, no.~12, pp. 6182--6207, 2010.

\bibitem{4215753}
B.~Kauffmann \emph{et~al.}, ``{Measurement-Based Self Organization of
  Interfering 802.11 Wireless Access Networks},'' in \emph{IEEE INFOCOM}, 2007,
  pp. 1451--1459.

\bibitem{borst2014nonconcave}
S.~C. Borst, M.~G. Markakis, and I.~Saniee, ``{Nonconcave utility maximization
  in locally coupled systems, with applications to wireless and wireline
  networks},'' \emph{IEEE/ACM Trans. Netw.}, vol.~22, no.~2, pp. 674--687,
  2014.

\bibitem{young93evolution-of-conventions}
H.~P. Young, ``{The Evolution of Conventions},'' \emph{Econometrica}, vol.~61,
  no.~1, pp. 57--84, 1993.

\bibitem{pradelski-young12nash-equilibria}
B.~S. Pradelski and H.~P. Young, ``{Learning Efficient Nash Equilibria in
  Distributed Systems },'' \emph{Games and Economic Behavior}, vol.~75, no.~2,
  pp. 882 -- 897, 2012.

\bibitem{Borowski2018}
H.~P. Borowski, J.~R. Marden, and J.~S. Shamma, ``Learning to play efficient
  coarse correlated equilibria,'' \emph{Dynamic Games and Applications}, Mar
  2018.

\bibitem{marden12state-based}
J.~R. Marden, ``{State Based Potential Games},'' \emph{Automatica}, vol.~48,
  no.~12, pp. 3075 -- 3088, 2012.

\bibitem{beherano-etal07fairness}
Y.~Bejerano, S.~J. Han, and L.~Li, ``{Fairness and Load Balancing in Wireless
  LANs Using Association Control},'' \emph{IEEE/ACM Transactions on
  Networking}, vol.~15, no.~3, pp. 560--573, June 2007.

\bibitem{coucheney-etal09user-association}
P.~Coucheney, C.~Touati, and B.~Gaujal, ``{Fair and Efficient User-Network
  Association Algorithm for Multi-Technology Wireless Networks},'' in
  \emph{IEEE INFOCOM 09'}, April 2009.

\bibitem{haddad-etal11hybrid-approach}
M.~Haddad, S.~E. Elayoubi, E.~Altman, and Z.~Altman, ``{A Hybrid Approach for
  Radio Resource Management in Heterogeneous Cognitive Networks},'' \emph{IEEE
  Journal on Selected Areas in Communications}, vol.~29, no.~4, pp. 831--842,
  April 2011.

\bibitem{singh-chaporkar13user-association}
M.~Singh and P.~Chaporkar, ``{An Efficient and Decentralised User Association
  Scheme for Multiple Technology Networks},'' in \emph{WiOpt 13', 11th Interna-
  tional Symposium on Modeling and Optimization in Mobile, Ad Hoc and Wireless
  Networks}, May 2013.

\bibitem{7746656}
S.~Ramakrishnan and V.~Ramaiyan, ``A completely uncoupled learning algorithm
  for general utility maximization,'' in \emph{2016 Int. Conf. on Signal
  Processing and Commun. (SPCOM)}, pp. 1--5.

\bibitem{doi:10.1137/110850694}
J.~Marden, H.~Young, and L.~Pao, ``Achieving pareto optimality through
  distributed learning,'' \emph{SIAM Journal on Control and Optimization},
  vol.~52, no.~5, pp. 2753--2770, 2014.

\bibitem{fan2018hoeffding}
J.~Fan, B.~Jiang, and Q.~Sun, ``Hoeffding's lemma for markov chains and its
  applications to statistical learning,'' \emph{arXiv preprint
  arXiv:1802.00211}, 2018.

\bibitem{isaacson1976markov}
D.~Isaacson and R.~Madsen, \emph{Markov chains, theory and applications}, ser.
  Wiley series in probability and mathematical statistics.\hskip 1em plus 0.5em
  minus 0.4em\relax Wiley, 1976.

\bibitem{8629359}
S.~{Ramakrishnan} and V.~{Ramaiyan}, ``Completely uncoupled algorithms for
  network utility maximization,'' \emph{IEEE/ACM Transactions on Networking},
  pp. 1--14, 2019.

\end{thebibliography}

\end{document}